\documentclass[
fleqn]{elsart}
\usepackage{amsmath,amssymb,dcolumn}

\def\be{\begin{equation}}
\def\ee{\end{equation}}
\def\bea{\begin{eqnarray}}
\def\eea{\end{eqnarray}}

\begin{document}
\begin{frontmatter}

\title{ On black hole spectroscopy via adiabatic invariance}
\author[a]{Qing-Quan Jiang\corauthref{cor}}
\ead{qqjiangphys@yeah.net}
\corauth[cor]{Corresponding author.} and
\author[b]{Yan Han}
\address[a]{\small College of Physics and Electronic Information, China
 West Normal University, Nanchong, Sichuan 637002, People's Republic of China}
\address[b]{\small College of Mathematic and Information, China
 West Normal University, Nanchong, Sichuan 637002, People's Republic of China}

\begin{abstract}
In this paper, we obtain the black hole spectroscopy by combining the black hole property of adiabaticity and the oscillating velocity of the black hole horizon. This velocity is obtained in the tunneling framework.
In particular, we declare, if
requiring canonical invariance, the adiabatic invariant quantity should be of the
covariant form $I_{\textrm{adia}}=\oint p_idq_i$. Using it, the horizon area of
a Schwarzschild black hole is quantized independent of the choice of coordinates, with an equally spaced spectroscopy always given
by $\Delta \mathcal{A}=8\pi l_p^2$ in the Schwarzschild and Painlev\'{e} coordinates.
\end{abstract}

\end{frontmatter}

\newpage

\section{Introduction}
The study of black hole physics has been a long-standing and hot topic in theoretical
physics since the birth of Einstein's theory of gravitation. In particular,
the exploration of the black hole entropy/area quantum has an important physical significance,
since it may provide a window to find an effective way to quantize a gravitational field.
However, a self-consistent theory of quantum gravitation was lacking so far. Hence, it
may be an appropriate juncture to ``take a step back" and reenforce our understanding of
these issues at a semiclassical level. The semiclassical notion that a black hole horizon
should be endowed with a quantum
area spectrum, has its origin traced back to the profound revelation of Bekenstein in the
early seventies \cite{Bekenstein1972,Bekenstein1973,Bekenstein1974}. The idea is based on the remarkable observation that the horizon area of a nonextremal black hole behaves as a classical \emph{adiabatic invariant}, which in the spirit of Ehrenfest principle corresponds to a quantum entity with discrete spectrum.
In 2002, Kunstatter \cite{r4} furthered this tapestry of idea to find, if a perturbed black hole is oscillating by the real part $\omega_R$ of the quasinormal mode frequencies,
the action $I=\int dE/\omega_R$ is an adiabatic invariant, which results in an equally spaced area spectrum with its value to be in agreement with the one given by Hod \cite{r3} as well as by
Bekenstein and Mukhanov \cite{r5}\footnote{For some work on this direction see \cite{Barvinsky2001,Bekenstein2002,Bekenstein200266,NEW2,NNEW2} and references therein.}. Here, one followed the Hod's proposal that the oscillating frequencies of a black hole only come from the real part of the highly damped qusinormal frequencies, and the imaginary part of it corresponds to the effective time for the black hole to return to a quiescent state \cite{r3}. In 2007, Maggiore refined the Hod's treatment by arguing that, insofar as a black
hole is to be viewed as a damped harmonic oscillator, the physically relevant frequency
would actually contain contributions from both real and imaginary parts of the complex quasinormal mode
frequencies \cite{r6}.
Combining this new interpretation
 for the black hole quasinormal frequencies with the black hole property of adiabaticity, Maggiore found the horizon area spectrum was exactly equal to the old result of Bekenstein  \cite{NEW1,NNEW1,NNEW3,NNEW4,Fernando79,Chen69,
Kwon28,Daghigh26,Kwon27,Kwon27165011,Kwon282011,Wei2010,Lopez-Ortega,MyungarXiv:1003.3519}.
 Recently, an interesting notion \cite{r7}\footnote{Also, there are some other methods proposed recently to investigate entropy spectrum and area spectrum \cite{rrr2,rrrr2,rrrr3,rrrr4,rrrr5,rrrr6}.} has shown that, utilizing solely the black hole property of
adiabaticity, one can also recover the Bekenstein's original result\footnote{ In fact, there is an error in \cite{r7} for studying the integration of the Euclidean time $\tau$. In Sec. \ref{sec3},
we will show that, if the adiabatic invariant quantity is of the form $\int p_idq_i$, as stated in \cite{r7},
a proper integration of the Euclidean time $\tau$ would give twice the result of Bekenstein. Only when it is of the form $\oint p_idq_i$, one
 can recover the result in \cite{r7}.}. It is noteworthy that there is no use at all of the quasinormal frequencies to obtain their findings. In fact, their ``so-called" interesting observation is lack of adequate consideration, which is reflected by: i) the quasinormal frequencies were really absent in \cite{r7} to obtain the black hole spectroscopy via adiabatic invariance, but the periodicity of the Euclidean time $\tau$ was introduced instead; ii) the adiabatic invariant quantity $\int p_idq_i$ proposed in \cite{r7} was not canonically invariant.

 In this letter, we combine the black hole property of adiabaticity with the oscillating velocity of the black hole horizon to revisit the black hole spectroscopy. As an exchange for no use of the quasinormal frequencies, we introduce the oscillating velocity of the black hole horizon to obtain the black hole spectroscopy via adiabatic invariance. In the tunneling framework, the black hole horizon can be assumed to oscillate periodically during the particle's tunneling process. This notion follows the Maggiore's proposal that the perturbed black hole can be treated as a set of harmonic oscillators. At this point, when a particle tunnels out or in, the action of the oscillating horizon is given by $I=\oint p_idq_i=\oint \frac{dH}{\dot{q}_i}dq_i$, where for the spherically symmetric horizon, $q_i$ corresponds to the Euclidean time $\tau$ and the black hole horizon $r_h$. Moreover, if one treats this action as an adiabatic invariant quantity, the black hole spectroscopy can be semiclassically fixed in the spirit of the Bohr-Sommerfeld quantization rule. Before that, it is necessary to first find the oscillating velocity of the black hole horizon $\dot{r}_h$. In the tunneling picture, when a particle tunnels out or in, the black hole horizon will oscillate due to the loss or gain of the black hole mass \cite{r8}. The tunneling is simultaneous with the oscillating. That is, the tunneling velocity of a particle is equal to the oscillating velocity of the black hole horizon $\dot{r}_h$ \cite{rrrr7}. With the aid of this oscillating velocity of the black hole horizon $\dot{r}_h$ and the adiabatic invariant quantity $I_{\textrm{adia}}=\oint p_idq_i$,
 we can quantize the horizon area of a Schwarzschild black hole.

The remainders of this paper are outlined as follows. In
Sec. \ref{sec1}, to verify the fact that the adiabatic invariant quantity $\int p_idq_i$ proposed in \cite{r7} is not canonically invariant, we quantize the black hole horizon in two different (Schwarzschild and Painlev\'{e} ) reference frames, and provide its corresponding area spectrum. Sec. \ref{sec2} is devoted to using the
adiabatic invariant quantity of the covariant form $I_{\textrm{adia}}=\oint p_idq_i$ to
revisit the black hole spectroscopy in the two reference frames. Sec. \ref{sec3} ends up with
some discussions and conclusions. Finally, a Kruskal extension in imaginary time appears in Appendix A.

\section{Adiabatic invariant quantity $I=\int p_idq_i$ and black hole spectroscopy} \label{sec1}

In \cite{r7}, it has been shown that when one uses $I=\int p_idq_i$ for the
adiabatic invariant quantity, one could recover the Bekenstein's area spectrum
by combining the black hole property of adiabaticity and the periodicity of the Euclidean time $\tau$.
In this section, we aim to examine this adiabatic invariant quantity in two different (Schwarzschild and Painlev\'{e} ) reference frames by combining the black hole property of adiabaticity and the oscillating velocity of the black hole horizon, and provide its corresponding black hole spectroscopy.

\subsection{Black hole spectroscopy in the Schwarzschild coordinate}\label{ssec1}

In the tunneling picture, when a particle tunnels out or in, the black hole horizon will be vibrated due to the loss or gain of the black hole mass \cite{r8}. According to \cite{r7}, the action of the oscillating horizon is given by
\begin{equation}
I=\int p_idq_i, \label{eq3}
\end{equation}
where $p_i$ is the conjugate momentum of the coordinate $q_i$. For the horizon of a Schwarzschild black hole, the coordinate $q_i$ corresponds to $q_0 = \tau$ and
$q_1= r_h$, where $\tau$ is the Euclidean time and $r_h$ is the horizon of the black hole. Applying the Hamilton's equation $\dot{q}_i=\frac{dH}{dp_i}$ where $H$ stands for the Hamiltonian of the system, this action is rewritten as
 \begin{equation}
\int p_idq_i=\int\int_0^H dH'd\tau+\int\int_0^H\frac{dH'}{\dot{r}_h}dr_h=2\int\int_0^H\frac{dH'}{\dot{r}_h}dr_h. \label{eq6}
\end{equation}
Here, $\dot{r}_h\equiv \frac{dr_h}{d\tau}$. Obviously, to evaluate the integral (\ref{eq6}), we must first find the oscillating velocity of the black hole horizon. In the tunneling picture, when a particle tunnels out or in, the black hole horizon will shrink or expand due to the loss or gain of the black hole mass. The tunneling and the oscillating take place at the same time. Naturally, the tunneling velocity of a particle $\dot{r}$ is equal and opposite to the oscillating velocity of the black hole horizon $\dot{r}_h$, that is \cite{rrrr7},
\begin{equation}
\dot{r}_h=-\dot{r}.   \label{eeq1}
\end{equation}
Specifically, for a Schwarzschild black hole,
 \begin{equation}
ds^2=f(r)d\tau^2+\frac{dr^2}{f(r)}+r^2d\Omega^2,  \label{eq5}
\end{equation}
where the metric is euclideanized by introducing the transformation $t\rightarrow -i\tau$, if a photon travels across the black hole horizon, the radial geodesics is given by
\begin{equation}
\dot{r}\equiv \frac{dr}{d\tau}=\pm i f(r),\label{eq7}
\end{equation}
where the $+(-)$ sign corresponds to the outgoing (ingoing) paths. When the photon tunnels out, the shrinking velocity of the black hole horizon is
\begin{equation}
\dot{r}_h=-\dot{r}=-i f(r).   \label{eq2}
\end{equation}
Then, the action (\ref{eq6}) is now read off
\begin{equation}
\int p_idq_i=-2i\int\int_0^H\frac{dH'}{f(r)}dr, \label{eq8}
\end{equation}
where we apply the relation $dr_h=-dr$. At the horizon, $r=2M$, there is a pole. To avoid it,
we take a contour integral over a small half-loop going above the pole from right to left. Now, if we treat
this action (\ref{eq8}) as an adiabatic invariant quantity, performing the integral as above yields \cite{rrrrr7,rrrrr8}
\begin{equation}
I_{\textrm{adia}}=\int p_idq_i=\pi \int_0^H\frac{dH'}{\kappa}, \label{eq9}
\end{equation}
Here, we only focus on the integration through $r=2M$
 since it is exactly where the adiabatic invariant quantity comes from\footnote{This
 is because the adiabatic invariant quantity, as a measurable one, is real.}, and $\kappa=\frac{1}{4M}$ is the surface gravity of the Schwarzschild black hole. It
is well-known that the black hole temperature is related to the surface gravity of the
horizon by $T=\frac{\hbar \kappa}{2\pi}$. Thus, the adiabatic invariant quantity can be rewritten as
\begin{equation}
I_{\textrm{adia}}=\int p_idq_i=\frac{\hbar}{2}\int_0^H\frac{dH'}{T}=\frac{\hbar}{2}S_{\textrm{bh}}, \label{eq10}
\end{equation}
where in the last step we have employed the first law of black hole thermodynamics $\int_0^H\frac{dH'}{T}=S_{\textrm{bh}}$. According
to the Bohr-Sommerfield quantization rule as described in \cite{r7}, the adiabatic invariant quantity has an equally spaced spectrum
in the semiclassical limit, that is
\begin{equation}
\int p_idq_i=nh.\label{neq10}
\end{equation}
Now, implementing this quantization rule (\ref{neq10}) for (\ref{eq10}), the black hole entropy is quantized evenly, with the spacing between the entropy spectrum given by
\begin{equation}
\Delta S_{\textrm{bh}}=4\pi. \label{eq11}
\end{equation}
Using the fact that the black hole entropy is proportional to the black hole horizon area, i.e. $S_{\textrm{bh}}=\frac{\mathcal{A}}{4l_p^2}$, the quantum of the horizon area is read off
\begin{equation}
\Delta\mathcal{A}=16\pi l_p^2. \label{eq12}
\end{equation}
Obviously, the horizon area is quantized and equidistant for the Schwarzschild black hole. Here, we have
used the action $\int p_idq_i$ for an adiabatic invariant quantity, to obtain the black hole spectroscopy in the Schwarzschild coordinate, and find a ``two" discrepancy between the result given by us (\ref{eq12}) and that in \cite{r7}. In fact, it will be shown in Sec. \ref{sec3} that there is an error for
investigating the integral of the Euclidean time $\tau$ in \cite{r7}, if it is appropriately
employed, the quantum of the horizon area (\ref{eq12})
can also be recovered.
Obviously, in the Schwarzschild frame, if one uses the action $\int p_idq_i$ for an adiabatic invariant quantity to fix the black hole spectroscopy, one would obtain twice the result of
Bekenstein. In the following subsection, we
examine this notion in Painlev\'{e} coordinate.

\subsection{Black hole spectroscopy in the Painlev\'{e} coordinate}\label{ssec2}

In this subsection, we continue to use $\int p_idq_i$ for the
adiabatic invariant quantity to obtain the black hole spectroscopy in the Painlev\'{e} coordinate. In
the Painlev\'{e} coordinate \cite{r9}, the coordinate singularity is removed at the black hole horizon by introducing a shifting of the time
coordinate, i.e.
\begin{equation}
d\tau'=d\tau+\frac{\sqrt{f(r)-1}}{f(r)}dr,  \label{eq13}
\end{equation}
for (\ref{eq5}), which yields
\begin{equation}
ds^2=f(r)d\tau'^2+2\sqrt{f(r)-1}d\tau' dr+dr^2+r^2d\Omega^2. \label{eq14}
\end{equation}
This is the Painlev\'{e}-Schwarzschild metric, in which there is not a coordinate singularity
any more at the black hole horizon, and constant-time slices are just flat Euclidean space.
These attractive features provide a superior setting for paths across the horizon. In this coordinate, when a particle tunnels out, the black hole horizon is shrinking by the velocity
\begin{equation}
\dot{r}_h= -i\big(1-\sqrt{1-f(r)}\big), \label{eq15}
\end{equation}
Now, proceeding in a similar procedure as in (\ref{eq9}), the adiabatic invariant quantity in Painlev\'{e} coordinate is read off
\begin{equation}
I_{\textrm{adia}}=\int p_idq_i=-2i\int\int_0^H\frac{dH'}{1-\sqrt{1-f(r)}}dr=2\pi\int_0^H\frac{dH'}{\kappa}. \label{eq16}
\end{equation}
Since the adiabatic invariant quantity corresponds to a quantum system with an equally spaced spectrum, according to
the Bohr-Sommerfield quantization rule (\ref{neq10}) we have
\begin{equation}
\Delta S_{\textrm{bh}}=2\pi, \label{eq17}
\end{equation}
which is the spacing level of the entropy spectrum of the Schwarzschild black hole. The same spacing was also obtained in \cite{rrr3,nrrr3,nrrr4}.
Obviously, in the Painlev\'{e} coordinate, the horizon area of the Schwarzschild black hole is quantized by
\begin{equation}
\Delta\mathcal{A}=8\pi l_p^2. \label{eq18}
\end{equation}
This result is identical to the Bekenstein's area quantum, but is half of the result (\ref{eq12})
derived in the Schwarzschild coordinate.
Obviously, the adiabatic invariant quantity of the form $I_{\textrm{adia}}=\int p_idq_i$ is
physically questionable since it is not invariant under coordinate transformations, and its resulting quantum of
the area spectrum apparently depends on the type of coordinates. How can one
reconcile these various results? In the next section, we propose, if the adiabatic invariant quantity of the covariant form $I_{\textrm{adia}}=\oint p_idq_i$, the area quantum is universally found independent of the
 choice of coordinates.

\section{Adiabatic invariant quantity $I_{\textrm{adia}}=\oint p_idq_i$ and black hole spectroscopy} \label{sec2}

In Sec. \ref{sec1}, the black hole spectroscopy is described by combining the black hole property of adiabaticity with the oscillating velocity of the black hole horizon. However, the proposed adiabatic invariant quantity $I_{\textrm{adia}}=\int p_idq_i$ apparently depends on the
choice of coordinates,
and one would obtain various results of the area quantum with change of coordinate transformations.
This is a physically questionable observation. In \cite{rrrrr7,rrrrr8}, it was argued that the closed contour
integral, $\oint p_idq_i$, was invariant under coordinate transformations, so we propose that the adiabatic invariant quantity should be of the covariant form $I_{\textrm{adia}}=\oint p_idq_i$. In this section, our effort is to clarify this proposal, and find the \emph{universal} quantum of the horizon area in the Schwarzschild and Painlev\'{e} coordinates.

\subsection{Black hole spectroscopy in the Schwarzschild coordinate}\label{sec2.1}
In the tunneling picture, during the particle's tunneling process, we assume the black hole horizon is vibrated periodically, with its action given by
\begin{equation}
I=\oint p_idq_i.\label{eq2.1}
\end{equation}
Near the horizon, the closed contour integral can be seen by considering
a closed path that goes from $q_i=q_{i}^{out}$, which is just outside the horizon, to $q_i=q_{i}^{in}$ just
inside the horizon, that is
\begin{equation}
\oint p_idq_i=\int_{q_{i}^{in}}^{q_{i}^{out}} p_i^{out}dq_i+\int_{q_{i}^{out}}^{q_{i}^{in}} p_i^{in}dq_i,\label{neq2.1}
\end{equation}
where $p_i^{out}$ ($p_i^{in}$) corresponds to the canonical momentum of the coordinates $q_i^{out}$ ($q_i^{in}$).
In the Schwarzschild coordinate, when a particle tunnels out or in a black hole, the oscillating velocity of the black hole horizon is
\begin{equation}
\dot{r}_h=\pm i f(r),
\end{equation}
where $+(-)$ sign corresponds to the expanding (shrinking) velocity of the black hole horizon. Applying the Hamilton equation, $\dot{q}_i=\frac{dH}{dp_i}$, we find the $p_i^{out}$ and $p_i^{in}$ have equal magnitude, but opposite signs. In this case, the closed contour integral is obtained by
\begin{equation}
\oint p_idq_i=2\int_{q_{i}^{out}}^{q_{i}^{in}} p_i^{in}dq_i=-4i\int_{r_{out}}^{r_{in}}\int_0^H\frac{dH'}{f(r)}dr.\label{neq2.2}
\end{equation}
Now, if we treats the action (\ref{eq2.1}) as an adiabatic invariant quantity, doing the contour integral as in (\ref{eq9}) for (\ref{neq2.2}) yields
\begin{equation}
I_{\textrm{adia}}=\oint p_idq_i=2\pi \int_0^H\frac{dH'}{\kappa}=\hbar \int_0^H\frac{dH'}{T}=\hbar S_{\textrm{bh}}.
\end{equation}
Here, the Hawking temperature of the black hole is given by $T=\frac{\hbar \kappa}{2\pi}$, and in
the last step we have exploited the first law of black hole thermodynamics
$\int_0^H\frac{dH'}{T}= S_{\textrm{bh}}$, where $S_{\textrm{bh}}$ is the black hole entropy.
Now, implementing the Bohr-Sommerfield quantization rule
\begin{equation}
\oint p_idq_i=nh,\label{nnnnneq}
\end{equation}
the spacing level of the entropy spectrum is given by
\begin{equation}
\Delta S_{\textrm{bh}}=2\pi.
\end{equation}
Thus, its resulting area quantum is read off
\begin{equation}
\Delta\mathcal{A}=8\pi l_p^2. \label{eqn1}
\end{equation}
This equally spaced area spectrum is half the result of (\ref{eq12}) obtained in the Schwarzschild frame by using the action $\int p_idq_i$ for an adiabatic invariant quantity, but in agreement
with the Bekenstein's original result. That is to say, when one uses the action $\oint p_idq_i$,
rather than $\int p_idq_i$ for an adiabatic
invariant quantity to quantize the black hole horizon in the Schwarzschild coordinate, the original result of Bekenstein could be well recovered. In the next subsection,
we remove this interest to the case of the Painlev\'{e} frame.

\subsection{Black hole spectroscopy in the Painlev\'{e} coordinate}

In the Painlev\'{e} frame, the coordinate is well behaved at the horizon by introducing a shifting
of time coordinate as in (\ref{eq13}).
In Sec. \ref{sec2.1}, we find, in the Schwarzschild coordinate, if one uses the action $\oint p_idq_i$ as an adiabatic invariant quantity, the spacing level of the area spectrum is in agreement with the original result by Bekenstein. In this subsection, we continue this
interest in the Painlev\'{e} coordinate. Here, the oscillating velocity of the black hole horizon is
\begin{equation}
\dot{r}_h= -i\big(\pm 1-\sqrt{1-f(r)}\big),
\end{equation}
where the $+(-)$ sign corresponds to the shrinking (expanding) velocity of the black hole horizon. Obviously, when doing the closed contour integral for the adiabatic invariant quantity $\oint p_idq_i$,
the integral $\int_{q_{i}^{in}}^{q_{i}^{out}} p_i^{out}dq_i$ has no contribution in the Painlev\'{e} coordinate, that is
\begin{equation}
\oint p_idq_i=\int_{q_{i}^{out}}^{q_{i}^{in}} p_i^{in}dq_i=-2i\int_{r_{out}}^{r_{in}}\int_0^H\frac{dH'}{1-\sqrt{1-f(r)}}dr.\label{eq3.2}
\end{equation}
Integrating (\ref{eq3.2}) near the black hole horizon, we have
\begin{equation}
I_{\textrm{adia}}=\oint p_idq_i=\hbar \int_0^H\frac{dH'}{T}=\hbar S_{\textrm{bh}}.
\end{equation}
According to the Bohr-Sommerfield quantization rule (\ref{nnnnneq}), this adiabatic system has an
equally spaced entropy spectrum in the semiclassical limit, that is
\begin{equation}
\Delta S_{\textrm{bh}}=2\pi,
\end{equation}
and an equally spaced area spectrum given by
\begin{equation}
\Delta\mathcal{A}=8\pi l_p^2.
\end{equation}
This result shows that, in the Painlev\'{e} frame where the coordinate is transformed from the Schwarzschild coordinate with a shifting of the time coordinate, if
one uses the action $\oint p_idq_i$ for an adiabatic invariant quantity, the black hole spectroscopy is quantized in
the same manner as that in the Schwarzschild coordinate. This is a physically desired result since the area quantum should be invariant under the coordinate transformations.
In a word, when studying the black hole spectroscopy via adiabatic invariance, if one treats the action $\oint p_idq_i$ as an adiabatic
invariant quantity, the black hole spectroscopy is
quantized independent of the choice of coordinates, and the Bekenstein's original result could be well recovered in the Schwarzschild and Painlev\'{e} coordinates.

\section{Conclusion and Discussion}\label{sec3}

In this paper, the black hole spectroscopy is intriguingly described by combining the black hole property of adiabaticity and the oscillating velocity of the black hole horizon. Unlike the Kunstatter's observation \cite{r4}, there is no use at all of quasinormal frequency, but the oscillating velocity of the black hole horizon has been introduced instead. To obtain the oscillating velocity, we cast this issue into the tunneling framework.
In particular, we declare, if
requiring invariance with change of the coordinate transformations, the adiabatic invariant quantity should be of the
covariant form $I_{\textrm{adia}}=\oint p_idq_i$. Using it, the black hole spectroscopy is quantized
independent of the choice of coordinates, and the Bekenstein's original result could be well recovered in
different types (Schwarzschild and Painlev\'{e}) of coordinates.

Some comments are followed: \textbf{\emph{i)}} In the isotropic frame, where the coordinates are transformed from the Schwarzschild coordinates with a shifting of the spatial variables, if one uses the action $\oint p_idq_i$ for an adiabatic invariant
quantity, one could also recover the Bekenstein's original result, as observed in the Schwarzschild and Painlev\'{e} coordinates.

\textbf{\emph{ii) }}In \cite{r7}, it was argued that the well-known result by Bekenstein could be well
recovered by using the action $\int p_idq_i$ for an adiabatic
invariant quantity. In Sec. \ref{ssec1} of this paper, we revisit the black hole
spectroscopy via this adiabatic invariant quantity, but find our derivation is twice the
result of Bekenstein. There is a ``two" discrepancy between them. Why appears such a difference? Next, we reconcile this
discrepancy by restudying the integration of the Euclidean time $\tau$ in \cite{r7}. In particular,
we will show our proposed adiabatic invariant quantity $I_{\textrm{adia}}=\oint p_idq_i$ is
universal and intriguing. In the Schwarzschild coordinate, there is a coordinate singularity at the horizon,
so it is necessary to use the Kruskal extension $(T, X)$ to connect the coordinates between the inside
and outside the horizon. When substituted by $T\rightarrow-i\mathcal{T}$ and $t\rightarrow-i\tau$, the Kruskal extension in imaginary time is given by (\ref{eqA1}).
 From this Kruskal-like extension it is found the
inside ($\mathcal{T}_{\textrm{in}}, X_{\textrm{in}}$) and outside ($\mathcal{T}_{\textrm{out}}, X_{\textrm{out}} $) coordinates
are connected with each other by the following relations\cite{rrrrr8}
\bea
\tau_{\textrm{in}}\rightarrow \tau_{\textrm{out}} -\frac{\pi}{2\kappa},~~~~~
(r_*)_{\textrm{in}}\rightarrow (r_*)_{\textrm{out}}-i\frac{\pi}{2\kappa}.
\eea
Now, proceeding in a similar way as in \cite{r7}, the
adiabatic invariant quantity is recalculated as $I_{\textrm{adia}}=2\int_{\tau_{\textrm{in}}}^{\tau_{\textrm{out}}}\int_0^H dH'd\tau=\pi\int_0^H \frac{dH'}{\kappa}$.
In view of the Bohr-Sommerfield quantization rule $\int p_idq_i=nh$ used in \cite{r7}, the quantum of the horizon area is obtained twice the result of Bekenstein, as given in Sec. \ref{ssec1}. Similarly, by
investigating the integration of the Euclidean time $\tau$ as above, we can also recover
the results in Sec. \ref{ssec2} for the black hole in the Painlev\'{e} coordinate. It
is noteworthy that, in the Painlev\'{e} coordinate, since the Euclidean time ($\tau'$) is related to that in the Schwarzschild
coordinate ($\tau$) via (\ref{eq13}), we have $\tau'_{\textrm{in}}\rightarrow \tau'_{\textrm{out}} -\frac{\pi}{\kappa}$. In a word,
when one uses the action $\int p_idq_i$ as an adiabatic invariant quantity, the quantum of the area spectrum always depends on the choice of coordinates.
On the other hand, when the adiabatic invariant quantity is taking the closed path of integration, i.e. $I_{\textrm{adia}}=\oint p_idq_i$, by investigating the
integration of the Euclidean time $\tau$, the Bekenstein's result would be well recovered independent of the choice of coordinates, as
observed in Sec. \ref{sec2}. Interestingly, we
once again verify the fact that, if requiring invariance with change of coordinate transformations, the adiabatic invariant
quantity should be of the form $I_{\textrm{adia}}=\oint p_idq_i$, rather than $I_{\textrm{adia}}=\int p_idq_i$.

\appendix

\section{The Kruskal extension in the imaginary time}
In the Schwarzschild coordinate, there is a coordinate singularity at the horizon,
so it is necessary to use the Kruskal extension
\bea
&&T_{\textrm{in}}=\frac{1}{\kappa}e^{\kappa (r_*)_{\textrm{in}}}\sinh \kappa t_{\textrm{in}},~~~~~~~X_{\textrm{in}}=\frac{1}{\kappa}e^{\kappa (r_*)_{\textrm{in}}}\cosh \kappa t_{\textrm{in}}, \nonumber\\
&&T_{\textrm{out}}=\frac{1}{\kappa}e^{\kappa (r_*)_{\textrm{out}}}\cosh \kappa t_{\textrm{out}}, ~~~X_{\textrm{out}}=\frac{1}{\kappa}e^{\kappa (r_*)_{\textrm{out}}}\sinh \kappa t_{\textrm{out}},
\eea
where $r_*=\int \frac{dr}{f(r)}$, to connect the coordinates between the inside
and outside the horizon. When substituted by $T\rightarrow-i\mathcal{T}$ and $t\rightarrow-i\tau$, we have
\bea
&&\mathcal{T}_{\textrm{in}}=\frac{i}{\kappa}e^{\kappa (r_*)_{\textrm{in}}}\cos \kappa \tau_{\textrm{in}},~~~~~~~X_{\textrm{in}}=-\frac{i}{\kappa}e^{\kappa (r_*)_{\textrm{in}}}\sin \kappa \tau_{\textrm{in}}, \nonumber\\
&&\mathcal{T}_{\textrm{out}}=\frac{1}{\kappa}e^{\kappa (r_*)_{\textrm{out}}}\sin \kappa \tau_{\textrm{out}}, ~~~X_{\textrm{out}}=\frac{1}{\kappa}e^{\kappa (r_*)_{\textrm{out}}}\cos \kappa \tau_{\textrm{out}}, \label{eqA1}
\eea
This is the Kruskal extension in the imaginary time.

\section*{Acknowledgments}
This work is supported by the National Natural Science Foundation of China with Grant No.
11005086, and by the Sichuan Youth Science and Technology
Foundation with Grant No. 2011JQ0019, and by a starting fund of China West Normal University with Grant No. 10B016.


\begin{thebibliography}{99}
\bibitem{r7}
B.R. Majhi and Elias C. Vagenas, Phys. Lett. B701, 623 (2011).

\bibitem{Bekenstein1972}
J. D. Bekenstein, Lett. Nuovo Cim. 4, 737 (1972).

\bibitem{Bekenstein1973}
J.D. Bekenstein, Phys. Rev. D 7, 2333 (1973).

\bibitem{Bekenstein1974}
J.D. Bekenstein, Lett. Nuovo Cimento 11, 467 (1974).

\bibitem{r4}
G. Kunstatter, Phys. Rev. Lett. 90, 161301 (2003) [gr-qc/0212014].

\bibitem{r3}
S. Hod, Phys. Rev. Lett. 81, 4293 (1998); Gen. Rel. Grav. 31, 1639 (1999).

\bibitem{r5}
J. D. Bekenstein and V. F. Mukhanov, Phys. Lett. B360, 7-12 (1995).

\bibitem{Barvinsky2001}
A. Barvinsky, S. Das, G. Kunstatter, Class. Quant. Grav. 18, 4845 (2001); Phys. Lett. B517, 415 (2001); Found. Phys. 32, 1851 (2002).

\bibitem {Bekenstein2002}
J. D. Bekenstein, Int. J. Mod. Phys. A17S1, 21 (2002).

\bibitem {Bekenstein200266}
J. D. Bekenstein, G. Gour, Phys. Rev. D 66, 024005 (2002).

\bibitem{NEW2}
 M. R. Setare, Phys. Rev. D 69, 044016 (2004); Gen. Rel. Grav. 37, 1411 (2005); Class.
Quant. Grav. 21, 1453 (2004).

\bibitem{NNEW2}
M. R. Setare and E. C. Vagenas, Mod. Phys. Lett. A 20, 1923 (2005).


\bibitem{r6}
M. Maggiore, Phys. Rev. Lett. 100, 141301 (2008).

\bibitem{NEW1}
E. C. Vagenas, JHEP 0811, 073 (2008).

\bibitem{NNEW1}
A. J. M. Medved, Class. Quant. Grav. 25, 205014 (2008).

\bibitem{NNEW3}
D. Kothawala, T. Padmanabhan, S. Sarkar, Phys. Rev. D 78 (2008) 104018.

\bibitem{NNEW4}
S.W. Wei, R. Li, Y.X. Liu, J.R. Ren, JHEP, 0903, 076 (2009).

\bibitem {Fernando79}
S. Fernando, Phys. Rev. D 79 (2009) 124026.

\bibitem {Chen69}
D. Y. Chen, H. T. Yang, X. T. Zu, Eur. Phys. J. C 69 (2010) 289-292.

\bibitem {Kwon28}
Y. Kwon,  S. Nam,  Class. Quantum Grav. 28 (2011) 035007.

\bibitem {Daghigh26}
R. G. Daghigh, M. D. Green, Class. Quantum Grav. 26 (2009) 125017.

\bibitem {Kwon27}
Y. Kwon,  S. Nam, Class. Quant. Grav. 27 (2010) 125007.

\bibitem {Kwon27165011}
Y. Kwon,  S. Nam, Class. Quant. Grav. 27 (2010) 165011,

\bibitem {Kwon282011}
Y. Kwon,  S. Nam, Class. Quantum Grav. 28 (2011) 035007

\bibitem {Wei2010}
S. W. Wei, Y. X. Liu, K. Yang, Y. Zhong, Phys. Rev. D 81 (2010) 104042.

\bibitem {Lopez-Ortega}
A. Lopez-Ortega, Class. Quantum Grav. 28 (2011) 035009.

\bibitem {MyungarXiv:1003.3519}
Y. S.   Myung,  Area   spectrum   of   slowly   rotating   black   holes, arXiv:1003.3519 [hep-th].

\bibitem{rrr2}
R. Banerjee, B. R. Majhi and E. C. Vagenas, Phys. Lett. B 686, 279 (2010).

\bibitem{rrrr2}
B. R. Majhi, Phys. Lett. B686, 49- (2010).

\bibitem{rrrr3}
R. Banerjee, B. R. Majhi, E. C. Vagenas, Europhys. Lett. 92, 20001 (2010).

\bibitem{rrrr4}
K. Ropotenko, Phys. Rev. D 82, 044037 (2010).

\bibitem{rrrr5}
Q.Q. Jiang, Y. Han, X. Cai, JHEP 1008, 049 (2010).

\bibitem{rrrr6}
Q.Q. Jiang, X. Cai, JHEP 1011, 066 (2010).

\bibitem{r8}
M.K. Parikh, F. Wilczek, Phys. Rev. Lett. 85, 5042 (2000).

\bibitem{rrrr7}
J.Y. Zhang, Z. Zhao, Phys. Rev. D83, 064028 (2011).

\bibitem{rrrrr7}
E.T. Akhmedov, V. Akhmedova, D. Singleton, Phys. Lett. B642, 124 (2006);

E.T. Akhmedov, V.Akhmedova, T. Pilling and D. Singleton, Int. J. Mod. Phys. A22 (2007) 1705; 

B. D. Chowdhury, Pramana 70 (2008) 593.

\bibitem{rrrrr8}
V. Akhmedova, T. Pilling, A. de Gill and D. Singleton, Phys. Lett. B666 (2008) 269; 

E.T. Akhmedov, T. Pilling and D. Singleton, Int. J. Mod. Phys. D17 (2008) 2453; 

V. Akhmedova, T. Pilling, A. de Gill and D. Singleton, Phys. Lett. B673 (2009) 227.

\bibitem{r9}
P. Painlev\'{e}, ``La m\'{e}canique classique et la th\'{e}orie de la relativit\'{e}," Compt. Rend. Acad. Sci. (Paris) 173, 677  (1921).

\bibitem{rrr3}
H. A. Kastrup, Phys. Lett. B385, 75 (1996).

\bibitem{nrrr3}
A. Barvinsky, G. Kunstatter, Phys. Lett. B389, 231 (1996).

\bibitem{nrrr4}
A. Barvinsky, G. Kunstatter, [gr-qc/9607030].





\end{thebibliography}
\end{document}